\documentstyle[11pt,epsf,MyM]{article}

\def\be{\begin{equation}}
\def\ee{\end{equation}}
\def\bea{\begin{eqnarray}}
\def\eea{\end{eqnarray}}

\begin{document}
\vspace*{4cm}
\title{CHIRAL DISORDER AND DIFFUSION IN THE QCD VACUUM}

\author{R.A. JANIK$^{1,2}$, 
\underline{M.A. NOWAK}$^{2,3}$, G. PAPP$^4$, I. ZAHED$^5$}

\address{
$^1$ Service de Physique Th\'{e}orique, CEA Saclay, F-91191
Gif-sur-Yvette, France.\\
$^2$ Department of Physics, Jagellonian University, 30-059 Krakow, Poland.\\
$^3$GSI, Planckstr. 1, D-64291 Darmstadt, Germany\\
$^4$ CNR Department of Physics, KSU, Kent, Ohio 44242, USA \& \\
HAS Research Group for Theoretical Physics, E\"{o}tv\"{o}s University,
     Budapest, Hungary\\
$^5$Department of Physics and Astronomy, Stony Brook, New York 11794, USA.%
}

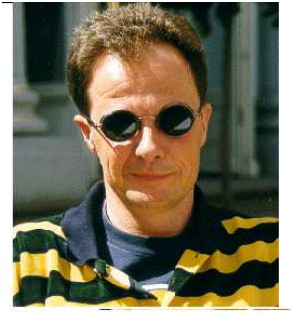
\maketitle\abstracts{
In this talk, which popularizes some of our recent work~\cite{PRLUS},
we provide novel insights into the bulk properties of light chiral
quarks in a fixed Euclidean volume ({\it e.g.} lattice QCD).
We show that the spontaneous breakdown of chiral symmetry 
results into diffusing quarks  with  a vacuum diffusion
constant $D=2F_{\pi}^2/|<\bar{q}q>|$  $\approx 0.22$ fm, in 
striking analogy to diffusing electrons in disordered metals
in one-, two- and  three-dimensions. }


The idea that light quarks diffuse in $D=4$ is key to understand
a number of phenomena in QCD in light of results known from 
disordered electronic systems. We introduce here the concept of 
the quark return probability in the QCD vacuum
as a chirally disordered medium~\cite{PRLUS}, borrowing from
concepts first introduced by Anderson~\cite{ANDERSON} in the context 
of localization. 

The eigenvalue equation for the  Euclidean Dirac operator 
for quarks in the fundamental representation and in the fixed
gluon field $A$ 
\bea
(i\nabla\!\!\!\!/[A]+im) q_k = \lambda_k[A]q_k 
\label{Dirac}
\eea
allows us to extend the theory into 4+1 dimensions with proper time
$\tau$, and to define the normalized 
return probability $P(\tau)$, for a light quark
to start at $x(0)$  and return back to the same position $x(\tau)$ after
a duration $\tau$ as,
\bea
P(\tau)=\frac{V^2}{N}\left< |\langle x(0)|x(\tau)\rangle|^2 \right>= 
\frac{V^2}{N}\left< |\langle
x(0)|e^{i(i\nabla\!\!\!\!/[A]+im)\tau}x(0)\rangle|^2 \right>\,.
\label{Evolution}
\eea
Here the Dirac operator acts as a four-dimensional Hamiltonian for the
evolution in proper time $\tau$, and $N$ is the mean number of quarks 
states in the four-volume $V=L^4$. The ensemble averaging is over the
gauge configurations which we note are $\tau$-independent (static
disorder). Using the arguments developed in~\cite{PRLUS}, we obtain
for large proper times
\bea
P(\tau)=e^{-2m\tau}\sum_{Q_{\mu}} e^{-DQ^2 \tau}\,,
 \label{diffusion}
\eea
with  $D=2F_{\pi}^2/|\langle\bar{q}q\rangle| \approx 0.22$ fm and
$Q_{\mu}=2\pi n_{\mu}/L$ with $n_{\mu}$ integers.
 In Fourier space $\lambda$ (dual to $\tau$),
\bea
P(\lambda)=\sum_Q\frac{1}{-i\lambda + \gamma + DQ^2}\,.
\label{Fourier}
\eea
The kernel of the sum in (\ref{Fourier}) corresponds to the kernel
of a classically diffusing particle in 4+1 dimensions. The damping factor
$\gamma=2m$ can be rewritten as $\gamma \equiv D/L^2_{coh}$, with the
help of the GOR relation~\cite{GOR}. $L_{coh}$ defines the coherence length for
light quarks, which is also the pion Compton wavelength $L_{coh}=1/m_{\pi}$.
In the chiral limit and for large times, the quark return probability
develops a diffusion pole at $\lambda=0$ and $Q^2=0$ (diffuson), which 
is a relic of the massless pion. 

It is now easy to identify all the relevant proper-time  scales
separating different regimes in the QCD vacuum viewed as a 
disordered medium. Because of the diffusion, we expect the emergence
of an ergodic time $\tau_{erg}=\sqrt{V}/D=L^2/D$,
which is the average time for the diffusing quarks to probe a sample 
of linear size $L$, as expected from Einstein's relation 
$\overline{x^2}(\tau) = D \tau$. For times greater than the ergodic time, 
but smaller than the Heisenberg time $\tau_h=1/\Delta$, where $\Delta$ is 
the average spacing between the quantum levels, we are in the ergodic 
(universal) regime dominated by the diffuson pole.
The return probability is then universal and  given by 
$P(\tau)=\exp (-2m \tau)$. By analogy with disordered
electronic systems, this universal regime should be described by 
random matrix theory.  For very large times (very small energies)
the classical (ergodic) description breaks down with the onset of a
purely quantum regime.

For times smaller than the ergodic time, the diffusing quark
probes only the part of the Euclidean volume $V$. 
All modes $Q_{\mu}$ are equally important,
and the return probability explicitly depends on the dimension of the
system.
This regime, called diffusive, ends at the elastic time $\tau_{elastic}=
1/2M_q$, where $M_q$ is the {\em constituent} mass of the quark.
For times shorter than the time to cross a mean-free path,
the concept of diffusion (and dressing of the quark through
multiple scatterings) becomes obsolete. The regime
$\tau  <\tau_{elastic}$ is referred to as ballistic.

In a  straightforward way we can  translate these scales onto the  
spectral properties of the Euclidean Dirac operator for 
finite volume QCD, {\it e.g.} lattice
calculations.  Using Heisenberg's uncertainty relation
$E \sim 1/t$ we define the spectral scale $\lambda \sim 1/\tau$.
The inverse of the ergodic time defines the Thouless virtuality 
$\Lambda_c=D/L^2=D/\sqrt{V}$, in analogy to the Thouless 
energy~\cite{THOULESS} in condensed  matter systems. 
The inverse of the Heisenberg time is just the mean quantum spacing 
between the Dirac eigenvalues  $\Delta$. Eigenvalues greater than $2M_q$ 
are part of the ballistic regime.

We depict various regimes of disorder in the spectrum of the 
Dirac operator in Fig.~1.

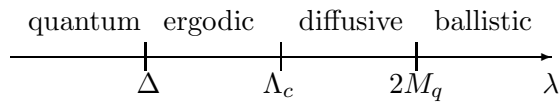
\begin{figure}[h]
\begin{center}
\setlength{\unitlength}{1.2mm}
\begin{picture}(60,10)
\put(0,4){\vector(1,0){60}}
\put(15,2.5){\line(0,1){3}}
\put(30,2.5){\line(0,1){3}}
\put(45,2.5){\line(0,1){3}}
\put(14,0){\mbox{$\Delta$}}
\put(28,0){\mbox{$\Lambda_c$}}
\put(42,0){\mbox{$2M_q$}}
\put(59,0){\mbox{$\lambda$}}
\put(2,7){quantum}
\put(17,7){ergodic}
\put(32,7){diffusive}
\put(47,7){ballistic}
\end{picture}
\end{center}
\caption{Disorder regimes in the eigenvalue 
spectrum of the Dirac operator.}
\end{figure}

The first milestone in this diagram is the average level spacing
$\Delta$ at small virtualities, which corresponds 
to the inverse of the spectral density at zero 
virtuality. It is directly related to the value of the quark condensate, 
via the Banks-Casher relation~\cite{BC}: 
$|\langle\bar{q}q\rangle|=\pi/(V\Delta )$.
 
The second milestone is played by the Thouless virtuality.
Here the crucial physical ingredient is the occurrence of the pion
decay constant through the diffusion constant. 

The third milestone is the breakdown of the diffusive picture,
corresponding to distance scales where the concept of dressing
a quark is meaningless. 

All these regimes are amenable to lattice verifications, as we explain
below. 

In the ergodic regime, the quark return probability is given solely
by the diffuson pole. Hence, any model with the same global symmetries 
as QCD, leading to $P(\tau) =e^{-2m\tau}$ would do for describing the
small quark eigenvalues. The simplest realizations
are chiral random matrix models in the microscopic limit~\cite{JACIS}, 
an outgrowth of the matrix models used in the macroscopic 
limit~\cite{USMAT,USTALK,SEMENOV}.
This regime is well accounted for by detailed lattice simulations~\cite{BB1}.
However, this agreement holds naturally only in the ergodic regime, 
where a direct relation to most of the dynamical observables in QCD is lost. 

In the diffusive regime, QCD is characterized by a complex dynamics
in D-dimensions. Nevertheless, there is still a possibility for a systematic 
study of the spectral properties, e.g. by semi-classical argument~\cite{PRLUS}
borrowing on arguments from mesoscopic systems~\cite{GUZ}. Indeed, the spectral
form factor~\cite{PRLUS,GUZ}
\bea
K(\tau) \approx \frac{2\tau \Delta^2}{4\pi^2\beta} P(\tau) \,,
\label{Bery}
\eea
is directly related to the average two-level spectral correlation function,
for different quark representations $\beta=1,2,4$, for pseudo-real, complex and
quaternion, respectively. In the diffusive regime, the spectral properties are 
reflecting on two-pion exchange, again in close analogy to the spectral 
properties of the two-diffuson and/or cooperon exchange
in disordered electron systems~\cite{ALTSCHULER}. 

Soon after our theoretical analysis, various numerical studies appeared
suggesting the occurrence of the Thouless virtuality. A numerical study
in the instanton model~\cite{OJ} and lattice QCD~\cite{BB2}, have 
confirmed some of our predictions, although the diffusion scenario
we have unraveled is yet to be explored. The first results reflecting 
on the spectral properties mentioned above are by now also 
available~\cite{GUHR}.

The ballistic regime is probably the most difficult one
to investigate on the lattice,
since the virtualities are large, and may interfere with 
the lattice cut-off.
We note, that a recent multi-matrix study~\cite{TAKA}  has 
confirmed that the ballistic regime is delineated by the constituent 
quark mass.

Last but not least, let us ask the crucial question about
the microscopic source of disorder in Euclidean QCD.
For disordered electron systems, the diffusion is triggered 
by the elastic scattering of electrons on external, {\em static}
defects in `dirty' wires, plaquettes or grains, for 
$D=1,2,3$  respectively. QCD is a fundamental theory, so where 
does the `dirt' come from? One possibility are instantons~\cite{NEGELE}. 
They are static quasi-particles in 4+1 dimensions, that cause a net 
chirality flip on light quarks (index theorem), with usually a random 
distribution in color and position space. However, other lumps of gauge-fields
can act similarly as well. In many ways, the renewal interest in random 
instanton systems~\cite{USMAT,DP,ED} can be regarded as the longstanding
motivation for the ideas developed in this note. We are pleased that early
suggestions that such systems are amenable to generic results from matrix
models~\cite{USMAT,USTALK,SEMENOV} and Anderson's ideas~\cite{USTALK} have come full 
circle, with potentially novel applications, some of which we have addressed
recently.

\section*{Acknowledgments}
M.A.N. is grateful to the organizers for the invitation
to such an enjoyable and productive meeting. 
This work was supported in part by the US DOE grants  DE-FG-88ER40388
and DE-FG02-86ER40251, by the 
Polish Government Project (KBN) grants 2P03B04412 and 2P03B00814 and by the 
Hungarian grant OTKA-F026622.

\end{document}